\documentclass[fleqn,usenatbib]{mnras}

\usepackage{newtxtext,newtxmath}
\usepackage[T1]{fontenc}
\usepackage{ae,aecompl}

\usepackage{graphicx}	% Including figure files
\usepackage{amsmath}	% Advanced maths commands
\usepackage{amssymb}	% Extra maths symbols

\title[The deepest \textit{Chandra} X-ray study of the plerionic supernova remnant G21.5$-$0.9]
{The deepest \textit{Chandra} X-ray study of the plerionic Supernova Remnant G21.5$-$0.9}
\author[B. T. Guest, S. Safi-Harb \& X. Tang] {
Benson T. Guest,$^{1}$\thanks{E-mail: umguest@myumanitoba.ca}
Samar Safi-Harb,$^{1}$\thanks{E-mail: samar.safi-harb@umanitoba.ca}
Xiaping Tang$^{2}$\thanks{E-mail: tangxiaping@gmail.com}
\\
$^{1}$Dept of Physics and Astronomy, University of Manitoba, Winnipeg, MB, R3T 2N2, Canada\\
$^{2}$The Racah Institute of Physics, The Hebrew University of Jerusalem, Jerusalem 91904, Israel\\
}

\date{Accepted XXX. Received YYY; in original form ZZZ}

\pubyear{2018}

\begin{document}
\label{firstpage}
\pagerange{\pageref{firstpage}--\pageref{lastpage}}
\maketitle

\begin{abstract}
G21.5$-$0.9 is a plerionic supernova remnant (SNR) used as a calibration target for the \textit{Chandra} X-ray telescope. The first observations found an extended halo surrounding the bright central pulsar wind nebula (PWN). A 2005 study discovered that this halo is limb-brightened and suggested the halo to be the missing SNR shell. In 2010 the spectrum of the limb-brightened shell was found to be dominated by non-thermal X-rays. In this study, we combine 15 years of \textit{Chandra} observations comprising over 1~Msec of exposure time (796.1~ks with the Advanced CCD Imaging Spectrometer (ACIS) and 306.1~ks with the High Resolution Camera (HRC)) to provide the deepest-to-date imaging and spectroscopic study. The emission from the limb is primarily non-thermal and is described by a power-law model with a photon index $\Gamma = 2.22 \, (2.04-2.34)$, plus a weak thermal component characterized by a temperature $kT = 0.37\, (0.20-0.64)$ keV and a low ionization timescale of \mbox{$n_{e}t < 2.95 \times 10^{10}$ cm$^{-3}$~s}. The northern knot located in the halo is best fitted with a two-component power-law + non-equilibrium ionization thermal model characterized by a temperature of 0.14 keV and an enhanced abundance of silicon, confirming its nature as ejecta. We revisit the spatially resolved spectral study of the PWN and find that its radial spectral profile can be explained by diffusion models. The best fit diffusion coefficient is $D \sim 2.1\times 10^{27}\rm cm^2/s$ assuming a magnetic field $B =130 \mu G$, which is consistent with recent 3D MHD simulation results.

\end{abstract}

\begin{keywords}
ISM: individual objects: G21.5--0.9 -- ISM: supernova remnants -- X-rays: ISM.
%keyword1 -- keyword2 -- keyword3
\end{keywords}

%%%%%%%%%%%%%%%%%%%%%%%%%%%%%%%%%%%%%%%%%%%%%%%%%%

%%%%%%%%%%%%%%%%% BODY OF PAPER %%%%%%%%%%%%%%%%%%

\section{Introduction}
 A pulsar wind nebula (PWN)\footnote{PWNe were originally referred to as `plerions' or `filled-center supernova remnants' as many of the original class were associated with known SNRs.  While most recently discovered PWNe have been found at X-ray to TeV energies, and have long outlived their natal SNR, the subclass of young PWNe still inside their SNR is an important one, of which G21.5-0.9 is an excellent example.  Hereafter we use
 %A PWN is historically referred to as `plerion' as it gives the supernova remnant (SNR) a centrally filled or `plerionic' morphology. Hereafter we use 
 the term `plerionic SNR' as it's commonly used in the literature when referring to SNRs hosting PWNe; see SNRcat~\url{http://www.physics.umanitoba.ca/snr/SNRcat} (\cite{Ferrand12}) for a list of all Galactic SNRs, including PWNe, and their classification.} 
 is a magnetic bubble of relativistic, non-thermal particles inflated by a rapidly rotating neutron star \citep{GS06}. It results from the interaction between the relativistic magnetized pulsar wind and the freely expanding supernova ejecta (or surrounding interstellar material). In the inner boundary of the PWN, the ram pressure of the pulsar wind is balanced with the confining pressure of the surrounding medium, leading to the formation of the so-called termination shock (TS). At the TS, relativistic particles injected by the rapidly rotating neutron star are thermalized and accelerated, which are then able to produce synchrotron emission from radio to X-ray energies  \citep[e.g.][]{GS06,Kargaltsev13}. At the outer boundary of the nebula, the magnetic field winds up and the shocked wind flow decelerates to match the velocity and pressure condition of the surrounding ejecta or circumstellar medium\footnote{A subclass of older PWNe is referred to as `bow shock nebulae' which form when the pulsar moves supersonically into the interstellar medium.}. As a result, a young PWN is centrally peaked in radio and X-ray emission. In radio, it is generally characterized by a flat spectral index.
 
G21.5$-$0.9 is a composite, young plerionic 
%supernova remnant 
SNR which displays a limb-brightened shell (\cite{Matheson-1}) and a central PWN. The PWN is significantly brighter than the surrounding shell and has been well studied at radio wavelengths. The initial mapping was done in the 1970's. \cite{becker+Kundu1976} and \cite{wilson+weiler1976} found a flat spectral index with $S_{\nu}\propto \nu^{-\alpha}, \alpha \sim 0.1$ and a strongly polarized elliptical brightness distribution which is peaked near the geometric centre, similar to the Crab Nebula and 3C~58. The first X-ray observations were taken with the Einstein Observatory. \cite{becker+szymkowiak1981} found the X-ray emitting region is comparable to the radio size. \cite{furst1988} suggested a symmetric double cone outflow structure, based on 22.3 GHz observations.
% and found that the X-ray maximum is located within a radio minimum suggesting the emission originates from different particle spectra.

The launch of \textit{Chandra} and \textit{XMM-Newton} in 1999 revealed X-ray structure beyond the PWN. A 150" radius halo was observed surrounding the PWN with knots of enhanced emission to the north (\cite{Slane-2000,Safi-Harb-2001,warwick2001}). The X-ray halo was found to have a non-thermal spectrum and was interpreted as either an extension of the PWN or a dust scattering halo (\cite{Safi-Harb-2001,Bocchino-2005}). There were problems with both of these interpretations. The X-ray PWN exceeding that seen in radio did not fit with evolutionary models and dust scattering alone could not account for the excess emission from the knots. \cite{bandiera2004} achieved a good fit to the diffuse emission with a dust scattering model and found evidence of a thermal component in the northern knot. \cite{Matheson-1} combined available \textit{Chandra} observations to reveal a candidate shell with limb-brightening observed at the eastern edge of the halo. The power-law photon index increased from the centre of the PWN to the edge, then remained flat within the halo out to a radius of 150$^{\prime\prime}$. \cite{Bocchino-2005} analysed \textit{Chandra} and \textit{XMM-Newton} data and interpreted the diffuse halo as a result of scattering off foreground dust
and the northern knot as shock-heated ejecta. \cite{Matheson-2} extended their previous work with additional \textit{Chandra} observations and found the knot required a thermal component supporting the ejecta assumption, while the limb was equally fit with non-thermal and thermal models. The non-thermal interpretation of the limb implied particle acceleration at the shock out to TeV energies. 
\cite{Matheson-2} also found the first evidence of variability in the PWN and weak thermal X-ray emission from the neutron star.

Radio pulsations from PSR J1833-1034 were detected independently by \cite{Gupta2005} and \cite{Camilo2006} who found a period of 62 ms, $\dot{P}=2.0 \times 10^{13}$, surface magnetic field of $B=3.6\times 10^{12}$ G, characteristic age of 4.8 kyr, and spin-down luminosity of $\dot{E}=3.3\times 10^{37}$ erg s$^{-1}$. \cite{Radio-Age} combined Very Large Array (VLA) observations from 1991 and 2006 to estimate a PWN expansion age of 870$^{+200}_{-150}$ yr\footnote{This age estimate neglects the expected acceleration that could increase the age by 20--25\%.}, which makes the remnant one of the youngest Galactic PWNe behind G29.7-0.3 with age estimates of about 400 years (\cite{Gelfand14,Reynolds18}). The shell remains undetected at radio wavelengths (\cite{B11}).
%add the above reference in the bib file
%http://adsabs.harvard.edu/abs/2011MNRAS.412.1221B

%In the hard X-ray and $\gamma$-ray observation, 
At higher energies, G21.5$-$0.9 has been detected by INTEGRAL in the soft $\gamma$-ray band \citep{Integral2004,Integral2009}, who found the PWN was dominant in the hard X-ray band, while H.E.S.S. observations showed the PSR J1833--1034 was the main source above 200 keV, with a 1--10 TeV flux of 2\% that of the Crab. \cite{Fermi-PSR2013} detected $\gamma$-ray pulsations with the Fermi Large Area Telescope (LAT).
\cite{Tsujimoto11} used G21.5$-$0.9 data to conduct a cross-calibration study of the instruments onboard \textit{Chandra}, \textit{Suzaku}, \textit{Swift},  \textit{XMM-Newton}, \textit{INTEGRAL} and \textit{RXTE}.  \cite{Nustar} used NuSTAR observations up to 40 keV and found evidence of the knot and limb features indicating non-thermal emission processes, with a break at 9 keV in the PWN spectrum. This break was most recently refined to 7.1~keV using \textit{Hitomi}'s broadband coverage (\citet{Hitomi-2018}). 

The distance to the SNR has been estimated by several authors to be in the 4.3--5.1 kpc range (e.g., \cite{Safi-Harb-2001,Camilo2006,Radio-Age,Kilpatrick-2016}).
%http://adsabs.harvard.edu/abs/2016ApJ...816....1K
A kinematic distance of 4.8 kpc was also proposed by \cite{Tian2008}. In this paper we adopt a distance of 5~kpc and refer to $D_5$ as the distance in units of 5~kpc.

\begin{figure}
	\includegraphics[width=\columnwidth]{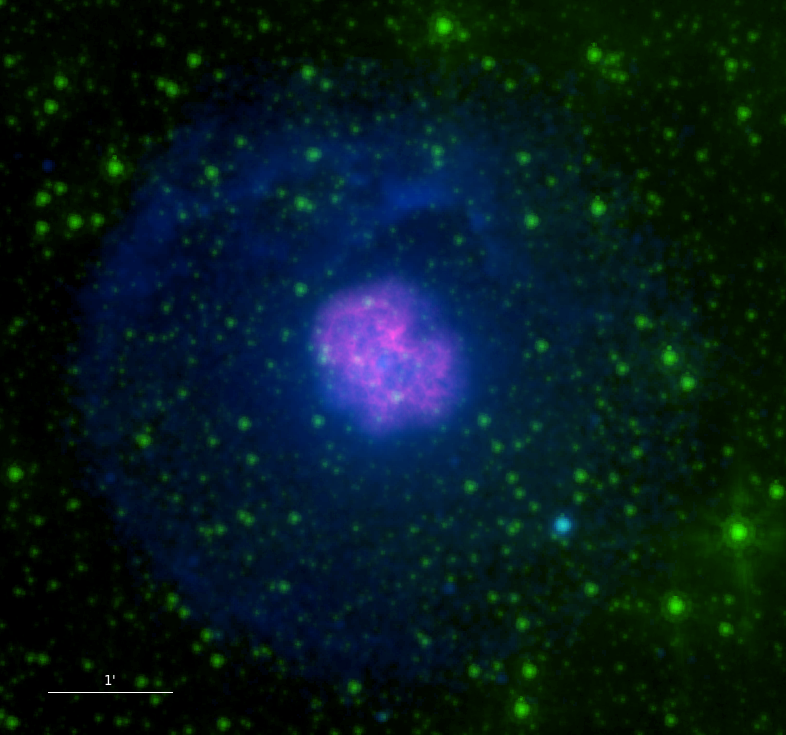}
	\caption{Multi-wavelength image of G21.5$-$0.9. From \textit{Chandra}: 0.5--10 keV (Blue), VLA: 4.75 GHz (Pink) from \protect\cite{Radio-Age}, Spitzer: 5.8 microns (Green) \protect\cite{Spitzer2012}. Note the 4.75 GHz data covers only the central PWN.}
	\label{FIG:RGB-Image}
\end{figure}

\section{Observations}
%The bright PWN dominates the emission of G21.5$-$0.9 and is an ideal candidate for a calibration target. 
As a calibration target for the  \textit{Chandra} X-ray observatory, G21.5--0.9 was observed regularly  by both the Advanced CCD Imaging Spectrometer (ACIS) and the High Resolution Camera (HRC). Recently, \cite{Matheson-2} analyzed 578.6 ks (1999-2006) of ACIS observations to study limb-brightening in the eastern limb and search for thermal emission. In this work, we extend this study utilizing 86 ACIS observations totaling of 786.1ks (1999-2014) and 17 HRC observations with totalling of 306.1ks. Data processing was performed using the \textit{Chandra} Interactive Analysis of Observations (\small{CIAO}) software package (\citet{CIAO}), while spectral analysis made use of the X-ray Spectral fitting package \small{XSPEC} version 12.9.1 (\citet{XSPEC}).

\subsection{Structure}
A multi-wavelength image is shown in Figure \ref{FIG:RGB-Image}. The remnant G21.5$-$0.9 is dominated by a bright PWN (visible from radio (pink) through x-ray (blue)), which is 40$^{\prime\prime}$ in radius, centered at $\alpha (2000) = 18^{h}33^{m}33.37^{s}$, $\delta (2000) = -10^{o}34'06".25$ and powered by the central pulsar PSR J1833--1034. The diffuse emission is only revealed by deep x-ray observations, extends to a radius of 150$^{\prime\prime}$ and displays a limb-brightened eastern edge. Knots with enhanced soft x-ray emission appear to the north and merge with the limb-brightened edge to the east. The diffuse emission fills a circle coincident with the geometry of the limb-brightened eastern edge, but the additional $\sim 200$ks of ACIS data still do not reveal limb-brightening to the west. The diffuse emission merely blends with the background level. Point source SS 397 is located to the south west within the extended shell. \cite{Spitzer2012} found infrared emission (green) from the inner 4$^{\prime\prime}$ but no extended emission at larger scales with the dominant contribution from field stars.

\subsection{Brightness Profile}
\begin{figure}
\includegraphics[width=\columnwidth]{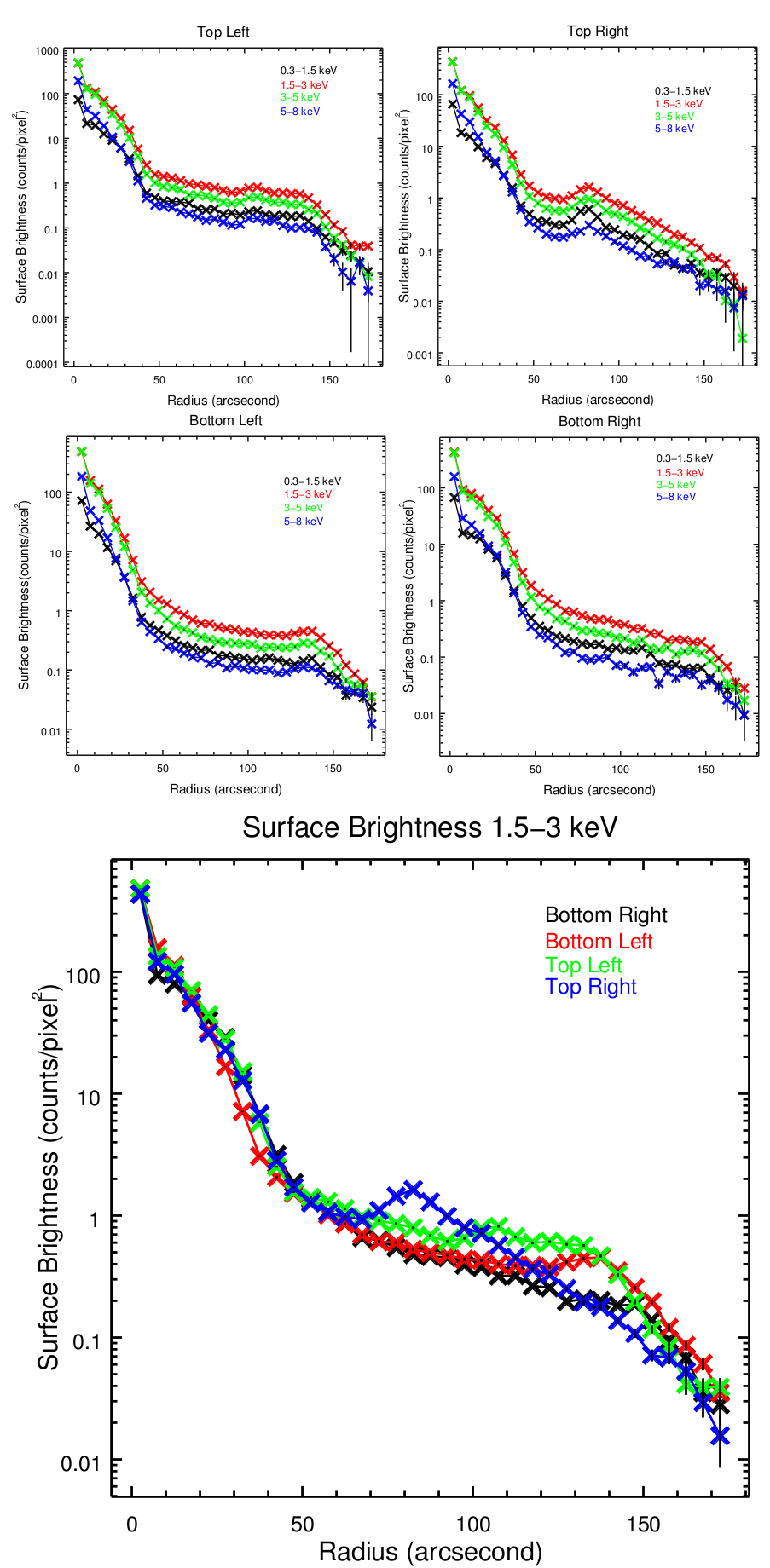}
\caption{Upper panel: Surface brightness profiles of four quadrants of G21.5-0.9 at four different energy bands. Bottom panel: Overlay of the profile of each quadrant in the 1.5--3 keV band.}
\label{FIG:SurfaceBrightness}
\end{figure}

The presence of a dust scattering halo complicates analysis of the faint outer regions identified as the supernova shell. We use the wealth of observations to update the surface brightness profiles of \cite{Bocchino-2005} and \cite{Matheson-1}. Profiles were extracted from quadrants of the merged event file and filtered into 4 energy bands: 0.3--1.5, 1.5--3, 3--5, and 5--8 keV. The results are shown in Figure \ref{FIG:SurfaceBrightness}. The bottom left quadrant shows the limb brightened edge revealed by \cite{Matheson-1}, the brightest knot is clearly visible in the top right profile. The top left shows the contribution from the fainter knot structures and the bottom right which displays no limb brightening smoothly declines to the background level.

\section{Spectroscopy}

\subsection{Processing}
Weighted spectra were extracted for each observation using the \small{CIAO} routine specextract. Each spectrum has an associated background which is extracted from the same CCD chip, but outside the shell and is not overlapping with any regions listed in the \textit{Chandra} Source Catalogue. The observations were processed with the \small{CIAO} 4.7 Chandra-repro script and is fitted simultaneously without merging the individual spectra as recommended by the \textit{Chandra} X-ray Center. The fitting was performed with the X-ray Spectral analysis software \small{XSPEC} v12.9.1 over the range 0.5--8 keV.

\subsection{Detector Contamination}
\begin{figure}
\includegraphics[width=\columnwidth]{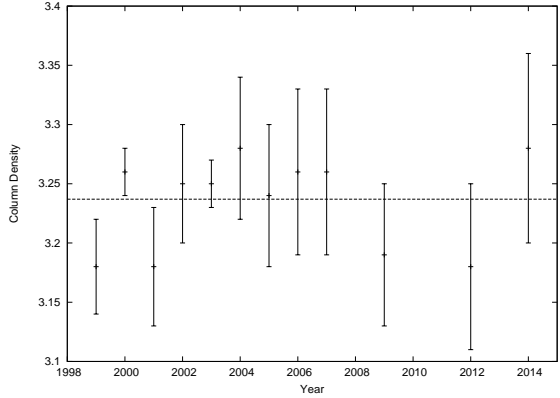}
\caption{Fitting observations of the PWN grouped by year to an absorbed power-law. The column density has units of $10^{22}$ atoms cm$^{-2}$. The dotted line marks the best fit value for all observations fit simultaneously. See \S3.3 for details.}
\label{FIG:ColumnDensity}
\end{figure}

Since the observational data are taken many years apart, we checked if the changing response affected our results. The build up of contaminants has a stronger effect on low energy emission \citep{Chandra-Contamination2004}, which is likely to affect the derived column density. To explore a systematic change in this parameter, we extract spectra from the PWN, group them by year of observation and fit individually to an absorbed power-law. We adopt the Tuebingen-Boulder ISM absorption model (\small{TBABS} in \small{XSPEC}), which calculates the cross-section for X-ray absorption. The required ISM abundances were set to those from \cite{TBABS-2000}. The results of this analysis are presented in Fig.\ref{FIG:ColumnDensity}. No general trend with increasing column density is found. Therefore, in what follows, we use the single best-fit value estimated with the total simultaneous fit in section \ref{PWN}.

\subsection{Pulsar Wind Nebula} \label{PWN}
In order to properly account for the absorption along the line of sight, spectra from the PWN were extracted from a 40$^{\prime\prime}$ circle centered at $\alpha (2000) = 18^{h}33^{m}33.37^{s}$, $\delta (2000) = -10^{o}34'06".25$. This was fit with an absorbed power-law with the absorption given by the \small{TBABS} model in \small{XSPEC} (see Table 1). The best fit parameter for the column density was frozen in subsequent region analysis to 3.237 $\times 10^{22}$ cm$^{-2}$ (see section \ref{S-RadProf} for further discussion).

\begin{table}
\caption{Spectral-fitting results for the pulsar wind nebula$^a$}
\label{TAB:pwn-fit}
\begin{tabular}{l c}
\hline \hline
$N_{H} (10^{22}$ atoms cm$^{-2})$ & 3.237 (3.225 - 3.250) \\
$\Gamma$ & 1.841 (1.835 - 1.847)\\
Norm ($10^{-2}$)$^{b}$ & 1.886 (1.869 - 1.903)\\
$ {\chi} ^{2}_{\nu}(\nu)$ & 1.114 (26369) \\
Absorbed Flux ($10^{-11}$ erg cm$^{-2}$ s$^{-1}$) & 4.555 $\pm$ 0.005\\
Luminosity($10^{35}$ erg s$^{-1}$)$^{c}$  & 2.822 $\pm$ 0.003 \\
Effective Exposure & 590.3 ks \\
\hline
\multicolumn{2}{ l }{$^{a}$ All confidence ranges are 90\%. Models were fit over} \\
\multicolumn{2}{ l }{the range 0.5-8 keV.} \\
\multicolumn{2}{ l }{$^{b}$ Units are photons keV$^{-1}$ cm$^{-2}$ s$^{-1}$} \\
\multicolumn{2}{ l }{$^{c}$ Unabsorbed, assuming a distance of 5 kpc to G21.5$-$0.9} \\
\end{tabular}
\end{table}

\subsection{Spectral Map}
\begin{figure}
\includegraphics[width=\columnwidth]{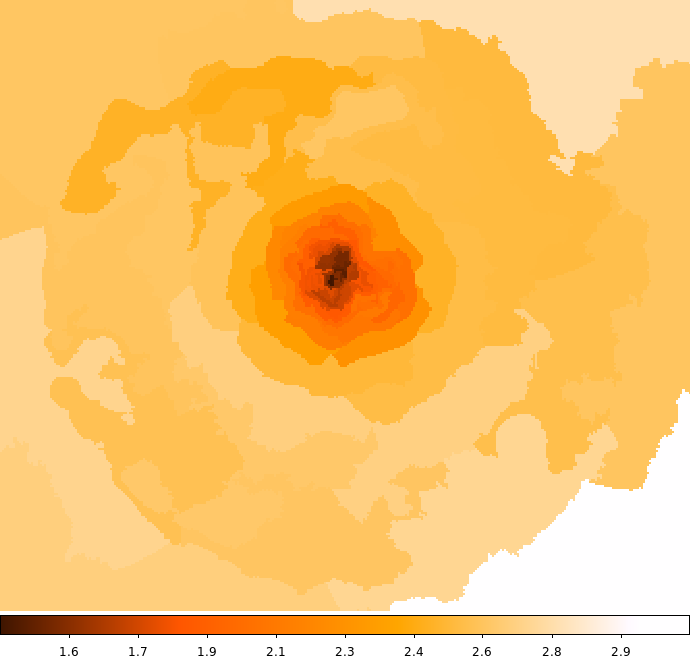}
\caption{Photon index map for the entire remnant. The colourbar shows the power-law photon index, with the darker (lighter) colour reflecting a harder (softer) spectrum.}
\label{FIG:SpectralMap}
\end{figure}

\begin{figure}
\includegraphics[width=\columnwidth]{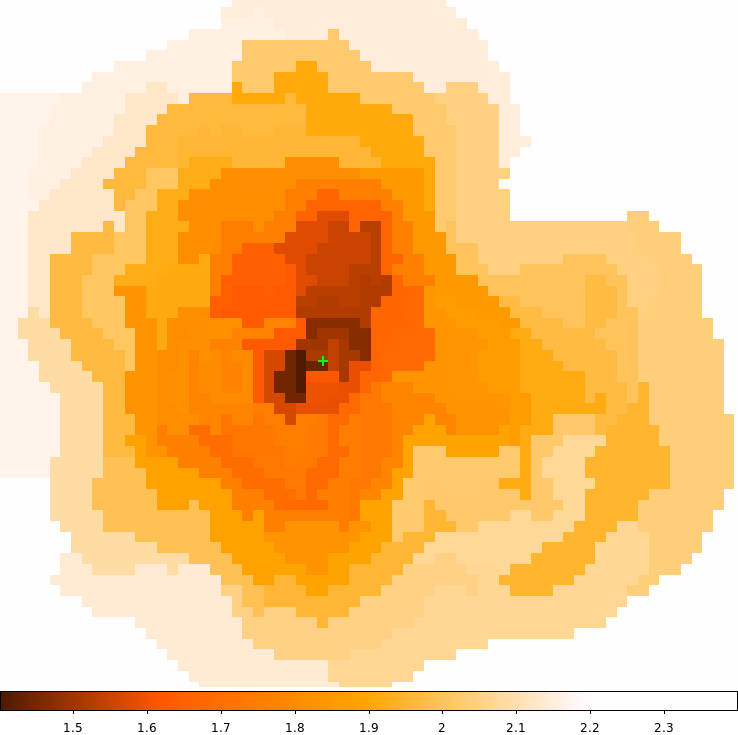}
\caption{Photon index map zoomed in on the PWN. The green cross marks the position of the PSR. See \S3.3.1 for details}
\label{FIG:SpectralMapZoom}
\end{figure}

To generate a photon index map, we applied the contour binning software \small{contbin}\footnote{http://www-xray.ast.cam.ac.uk/papers/contbin/} \citep{contbin}, which produces adaptive bin size following the surface brightness of an input image, such that each bin meets a given signal limit.
The exposure corrected merged flux counts image spanning the range 0.5--10 keV was used as input with a signal limit of 150. This corresponds to a limit of a few hundred counts in the resulting spectra for a single observation of each generated region. The spectra were fitted with an absorbed power-law and the absorption coefficient is frozen to the best fit value derived for the PWN (see section \ref{PWN}). The generated regions were coloured with their best fit value of the photon index, which is illustrated in Fig. \ref{FIG:SpectralMap}.  A zoomed in view of the small scale in the PWN is presented in Figure \ref{FIG:SpectralMapZoom} with details. The non-symmetric nature of the emission is clearly shown in the figure, where the hardest emission is offset from the location of the PSR (marked with a cross) with a bubble of higher energy emission extending to the north. 

\subsection{Radial Profile}\label{S-RadProf}
\begin{figure}
	\includegraphics[width=\columnwidth]{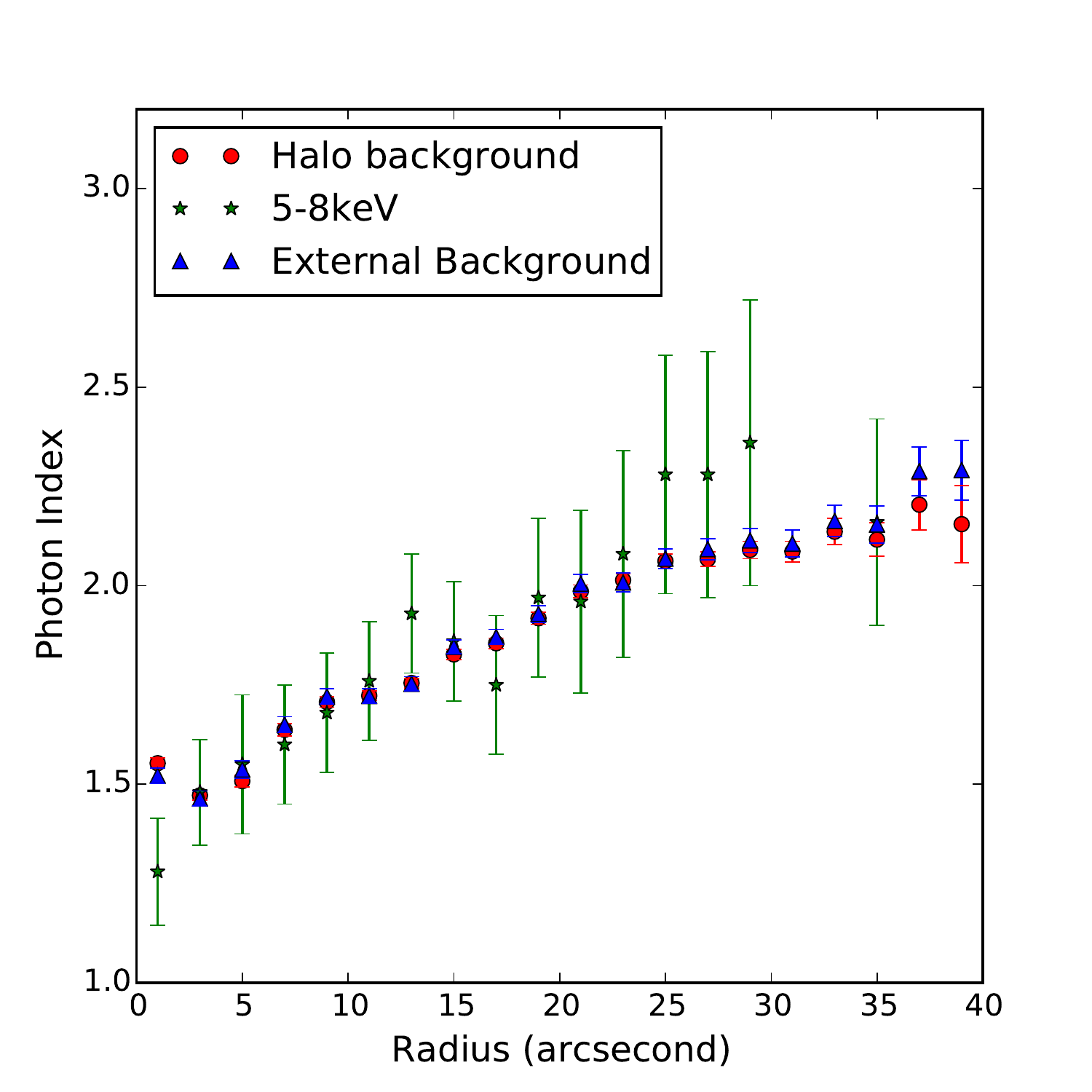}
	\caption{Photon index found for rings of increasing size within the central bright PWN. The External (triangles, blue symbols) and Halo (circles, red symbols) background profiles correspond to the 0.5--8 keV bands and use a background region outside and from within the halo, respectively. The 5--8 keV profile (stars, green points)  also use the internal background, which is
    %blue points use a background region outside of the shell, while the red and green points use 
    an annulus centred on the pulsar with a radius of 44$^{\prime\prime}$--48$^{\prime\prime}$ to minimize contribution from the dust scattering halo. Errorbars on the photon index are at the 90\% confidence level.}
	\label{FIG:PWNRadialProfile}
\end{figure}

Spectra were extracted from rings centered on the pulsar, and fit with an absorbed power-law. The column density is found to be $3.237 \times 10^{22} \, cm^{-2}$ for the external background region selection and $3.222\times 10^{22} \, cm^{-2}$ for the internal background selection. As shown in Fig. \ref{FIG:PWNRadialProfile}, the photon index is shown to increase to the edge of the PWN at 40$^{\prime\prime}$, which is consistent with previous studies (e.g. Matheson \& Safi-Harb 2005).  However, a higher column density is derived comparing with previous work (Slane et al. 2000,  Safi-Harb et al. 2001,  Warwick et al. 2001, Bocchino et al. 2005,  Matheson \& Safi-Harb 2010) , which is likely because we apply the \small{TBABS} model with $wilm$ abundances. We verify this by fitting the PWN spectra with the previously used \small{WABS} model for photoelectric absorption which uses the Wisconsin cross-sections (\cite{WABS}) and find a smaller value of $2.37\pm 0.01 \times 10^{22}$cm$^{-2}$ which is consistent with previously published results. To account for the halo emission, an annulus centred on the pulsar with radius 44$^{\prime\prime}$--48$^{\prime\prime}$ was chosen as background which had a negligible effect on the spectral index. This is expected due to the drop in surface brightness outside the PWN. Figure \ref{FIG:SurfaceBrightness} shows that the surface brightness where the internal background was taken is fainter by a factor of more than 20 from the inner PWN. Therefore the dust scattered halo component leads to a systematic underestimation of the photon index errors, yet the overall result will remain unchanged. Additionally, the halo contributes minimally above 5 keV (\cite{Bocchino-2005}) and when we restrict our fits to the range 5--8 keV we find the trend is unchanged. The high energy profile differs in that the pulsar component is not visible in the first data-point.

\begin{figure}
\includegraphics[width=\columnwidth]{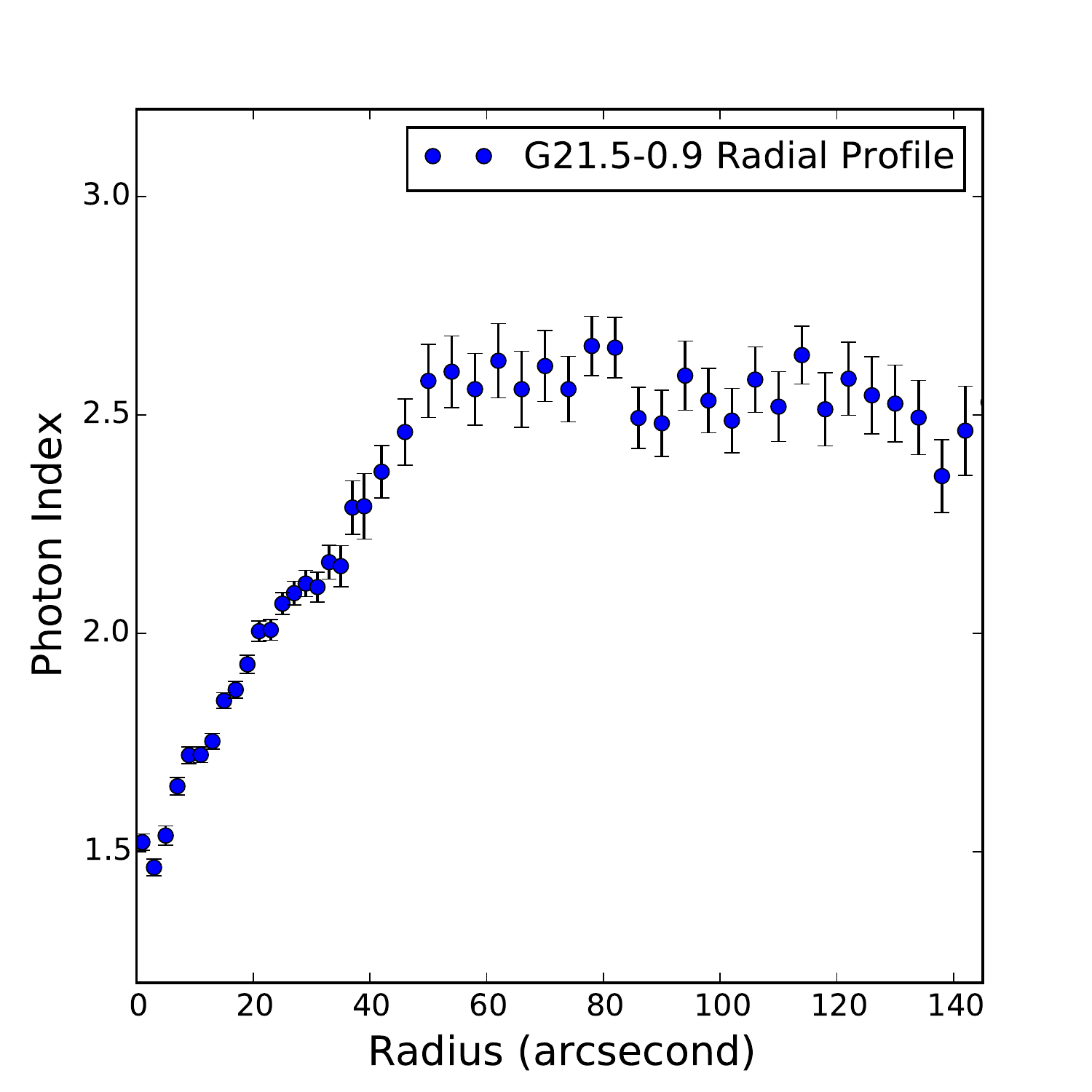}
\caption{Spectral index vs radius fit for the entire remnant over 0.5--8 keV, with the absorption frozen to the best fit value found from fitting the PWN. The northern knot region coincides with the slight increase at 80$^{\prime\prime}$ and the limb-brightened edge of the remnant corresponds to the drop at 140$^{\prime\prime}$.}
\label{FIG:RemnantRadialProfile}
\end{figure}

The near linear increase of the spectral index with radius is consistent with that observed in other young PWNe, such as 3C~58 (\cite{Slane04}). The spectral index continues to rise beyond the edge of the PWN at 40$^{\prime\prime}$, which reaches a maximum at 50$^{\prime\prime}$ and remains roughly flat to the edge of the remnant, see Figure \ref{FIG:RemnantRadialProfile}. The model fitting of the spectral index profile will be discussed in detail in section \ref{sec:PWN_fitting}. We include the best fit power-law parameters in the appendix to assist with future modeling work (see Tables A1-A4), and in spatially resolved spectroscopic studies of the whole remnant, including the halo and shell emission.

\subsection{PSR J1833--1034}
\begin{figure}
	\includegraphics[width=\columnwidth]{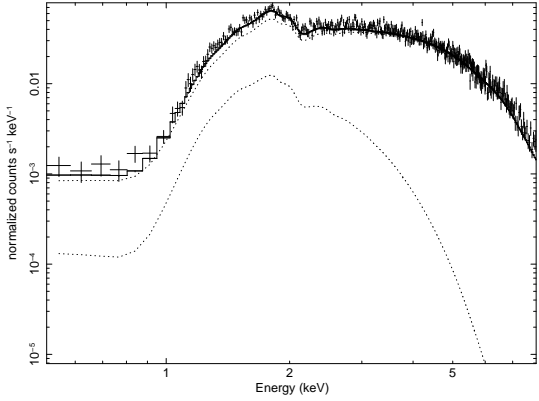}
	\caption{Combined spectra of the PSR with the simultaneous model fit (solid line), power-law and blackbody components (dashed lines).}
	\label{FIG:PSR-fit}
\end{figure}

Evidence of weak thermal x-ray emission from the pulsar powering $G21.5-0.9$ has been suggested
%90\% of the energy of a 1.5 keV point source is contained within a radius of 1$^{\prime\prime}$.5 at an off axis angle of 3$^{\prime}$, which is suggested as the thermal emission from the surface of the neutron star 
(Matheson \& Safi-Harb 2010). To follow up on this study with the deeper exposure,
we select a 2$^{\prime\prime}$ radius centred on the pulsar at $\alpha (2000) = 18^{h}33^{m}33.54^{s}$, $\delta (2000) = -10^{o}34'07.6"$ using observations with an off-axis angle less than 3$^{\prime}$\footnote{90\% of the energy of a 1.5 keV point source is contained within a radius of 1$^{\prime\prime}$.5 at an off axis angle of 3$^{\prime}$}. The background was extracted from a circular annulus centred on the PSR with radius 2$^{\prime\prime}$-- 4$^{\prime\prime}$ to remove contamination from the PWN. The spectra were fitted with an absorbed power-law utilizing the best fit column density in the PWN. The fitting results are provided in Table \ref{TAB:psr-fit}, and a combined spectrum is shown in Figure \ref{FIG:PSR-fit}. The single power-law model derives a hard spectral index $\Gamma =1.54 \pm 0.02$ with $\chi^{2}_{\nu}(\nu) = 1.075 (3731)$. The addition of a thermal blackbody component improves the fit slightly, which yields a harder spectral index $\Gamma = 1.35 \pm 0.12$ with temperature $kT = 0.43^{+0.04}_{-0.09}$ keV and $\chi^{2}_{\nu}(\nu) = 1.072 (3729)$. F-test is a statistical measure of the requirement of an additional model based on improvement in the reduced chi-squared value vs the change in the number of degrees of freedom. We find that the thermal component is required over a power-law alone with an F-test probability of 7.6$\times 10^{-3}$. This is notably less significant than the previous result by \cite{Matheson-2}, which found a probability of $2.6 \times 10^{-4}$.

\begin{table}
\caption{Spectral-fitting results for PSR J1833--1034.}
\label{TAB:psr-fit}
\begin{tabular}{c c c}
\hline \hline

Model & Model Parameter & PSR \\
\hline
 & Effective Exposure & 280.2 ks \\
\hline
Power-law & $N_{H} (10^{22}$ atoms cm$^{-2})$ & 3.237 (Frozen) \\
 & $\Gamma$ & 1.54 (1.52 - 1.56) \\
 & Norm ($(10^{-4})^a)$ & 8.34 (8.13 - 8.57) \\
 & $ {\chi} ^{2}_{\nu}(\nu)$ & 1.075 (3731) \\
 & Flux ($10^{-12}$)$^{b}$ & 3.19 $\pm$ 0.02 \\
\hline
Power-law & $N_{H} (10^{22}$ atoms cm$^{-2})$ & 3.237 (Frozen) \\
+ & $\Gamma$ & 1.35 (1.23 - 1.46) \\
Blackbody & Norm ($(10^{-4})^a)$ & 6.14 (5.04 - 7.08) \\
 & kT (keV) & 0.43 (0.34 - 0.47) \\
 & Norm ($(10^{-6})^c)$ & 5.74 (2.89 - 8.68) \\
 & $ {\chi} ^{2}_{\nu}(\nu)$ & 1.072 (1421) \\
 & Flux ($10^{-12}$)$^{b}$ & 3.16 (3.13 - 3.18) \\
 & Thermal flux($10^{-13}$)$^{b}$ & 1.18 \\
 & Non-thermal flux  ($10^{-12}$)$^{b}$ & 3.04 \\
 & Thermal flux ($10^{-13}$)$^d$ & 4.57 \\
\hline

\multicolumn{3}{ l }{$^{a}$ photons keV$^{-1}$ cm$^{-2}$ s$^{-1}$} \\
\multicolumn{3}{ l }{$^{b}$ Observed flux in units of erg cm$^{-2}$ s$^{-1}$} \\
\multicolumn{3}{ l }{$^{c}$ $L_{39}/D^{2}_{10}$ }\\
\multicolumn{3}{ l }{$^d$ Unabsorbed flux in units of erg cm$^{-2}$ s$^{-1}$}\\
%\multicolumn{3}{ l }{where $L_{39}=L/(10^{39}$ erg s$^{-1})$ and $D_{10}=D/(10 $ kpc)} \\
\end{tabular}
\end{table}

\subsection{Northern Knot}\label{S-Knot}
\begin{figure}
	\includegraphics[width=\columnwidth]{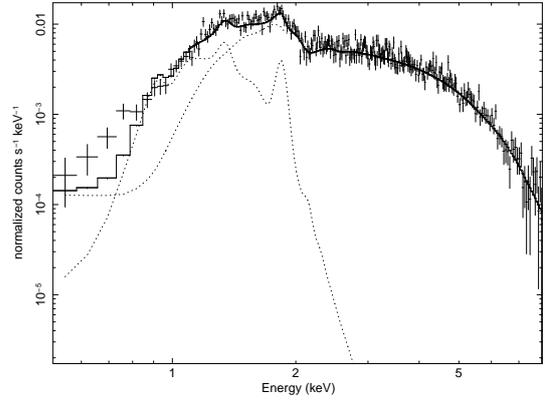}
	\caption{Combined Spectra of the knot plotted with the simultaneous fit result for a power-law + \small{VPSHOCK} model. See text for details.}
	\label{FIG:Knot-fit}
\end{figure}

The bright knot to the north of the PWN appears as a region of enhanced soft X-ray emission. Previous studies suggested a two-component model with both thermal and non-thermal emission. However, the thermal emission component was not well constrained. \cite{Bocchino-2005} found evidence for thermal emission in the `North Spur', which is characterized by a two-component power-law plus vnei model. The fitting to \textit{Chandra} and \textit{XMM-Newton} data produced two minima: one is consistent with solar abundances and the other with enhanced abundance of Si (2--20 times solar) and Mg (0.6--3 times solar). \cite{Matheson-2} show the northern knot is dominated by non-thermal emission while the thermal component comprising only 6\% of the observed flux. The abundances are more consistent with the solar abundances provided by Bocchino et al. with $Mg$ = 0.72 (0.40--1.06), $Si$ = 0.84 (0.32--1.35)  although also included a large (albeit poorly constrained) abundance of Sulphur, $S$ = 107 (4--210).
In addition to the two-component model, we fit the data with thermal and non-thermal models separately. The results (see Table \ref{TAB:Knot-fit} and Figure \ref{FIG:Knot-fit}) are consistent with the previous \textit{Chandra} study. The thermal model requires solar abundances but a higher temperature. The addition of a non-thermal component improves the fit, meanwhile lowering the required temperature and enhancing the abundance of Si. The abundance of Si is poorly constrained and tied to the \small{VPSHOCK} normalization. When both parameters are allowed to vary, there is no plausible upper limit found on the Si abundance; however the lower limit is still characteristic of enhancement. In order to place plausible constraints on the abundance, the normalization of the \small{VPSHOCK} model was frozen at its best fit value for the calculation of the Silicon error range.
Our results are consistent with the 2nd minima discussed in \cite{Bocchino-2005} and supports the interpretation of an ejecta knot of Si. Assuming the emitting volume is an ellipsoid with $V = 1.77\times 10^{55}D_{5}f cm^{3}$, where $f$ is the filling factor, we estimate the density of the emitting plasma to be $42$~cm$^{-3}$, which is significantly higher than the ambient density calculated in section \ref{S-Limb}.

\begin{table}
\caption{Spectral-fitting results for the northern knot. See text for details.}
\label{TAB:Knot-fit}
\begin{tabular}{c c c}
\hline
\hline
Model & Parameter & Northern Knot \\
\hline
& Effective Exposure  & 509.4 ks \\
&  $N_{H} (10^{22}$ atoms cm$^{-2})$ & 3.237 (Frozen) \\
\hline
Power-law & $\Gamma$ & 2.62 (2.578-2.67) \\
 & Norm ($10^{-4}$)$^a$ & 3.67 (3.51-3.84) \\
 &  $ {\chi} ^{2}_{\nu}(\nu)$ & 1.306 (815) \\
 & Flux ($10^{-13}$)$^{b}$ & 3.12 (3.07-3.16) \\
\hline
vpshock & kT (keV) & 4.2 (3.8-4.7) \\
 & Mg & 0.56 (0.39-0.72) \\
 & Si & 0.31 (0.24-0.39) \\
 & S  & 0.16 (0.04-0.28)\\
 & $n_{et} (10^{10}$cm$^{-3}$ s) & 2.86 (1.70-4.75) \\
 & Norm ($10^{-4}$ cm$^{-5}$) & 5.04 (4.64-5.55)\\
 &  $ {\chi} ^{2}_{\nu}(\nu)$ & 1.062 (811) \\
& Flux ($10^{-13}$)$^{b}$ & 3.33 (3.27-3.36) \\
\hline
Power-law& $\Gamma$ & 2.24 (2.17-2.31) \\ %Updated July 14
+ & Norm ($10^{-4}$)$^a$ & 2.51 (2.28-2.72) \\
vpshock & kT (keV) & 0.15 (0.12-0.17) \\
& Mg & 0.77 (0.35-1.72) \\
& Si & 15.7 (7.7-26.2) \\
& $n_{et}$($10^{13}$ cm$^3$ s$^{-1}$) & 5.0 ($>$0.3) \\
& Norm (cm$^{-5}$) & 0.102 (0.042-0.177) \\
& $ {\chi} ^{2}_{\nu}(\nu)$ & 1.041 (1178) \\
& Flux ($10^{-13}$)$^{b}$ & 3.20 (3.04-3.21) \\
& Non-thermal flux ($10^{-13}$)$^{b}$ & 3.05 \\
& Thermal flux ($10^{-13}$)$^{b}$ & 0.15 \\
\hline
Power-law$^{c}$ & $\Gamma$ & 2.55 (Frozen) \\ 
+ & Norm ($10^{-4}$)$^a$ & 2.70 (1.64-2.96) \\
Power-law& $\Gamma$ & 1.3 (0.4-2.1) \\ 
+ & Norm ($10^{-5}$)$^a$ & 2.0 (1.6-8.0) \\
vpshock & kT (keV) & 0.14 (0.13-0.15) \\
& Mg & 1.1 (0.5-1.8) \\
& Si & 37 (5-118) \\
& $n_{et}$($10^{13}$ cm$^3$ s$^{-1}$) & 2.7 ($>$0.04) \\
& Norm (cm$^{-5}$) & 0.14 (0.08-0.28) \\
& $ {\chi} ^{2}_{\nu}(\nu)$ & 1.041 (1176) \\
\hline
\multicolumn{3}{ l }{$^{a}$ Photons keV$^{-1}$ cm$^{-2}$ s$^{-1}$} \\
\multicolumn{3}{ l }{$^{b}$ Observed flux in erg cm$^{-2}$ s$^{-1}$}\\
\multicolumn{3}{ l }{$^{c}$ Dust scattering halo component}
\end{tabular}
\end{table}

The dust scattering halo complicates the analysis of the faint thermal emission. To better understand its effect on our fits we extract a background from an annulus surrounding the knot. The general result does not change. We find a power-law photon index and temperature that are consistent (within error) with the external background. The abundances show the same trend with an approximately solar abundance of Mg and an enhanced abundance of Si; however the parameters are less bound with the upper limit for Si remaining unconstrained.
Alternatively, we try adding an additional power-law component with its photon index fixed to the value at which the radial profile (see Figure \ref{FIG:RemnantRadialProfile}) appears to level off, and allow the normalization to vary. Again, the thermal parameters are consistent with the previous result. A harder photon index of 1.3 is found for the non-thermal component yet this addition does not improve the fitting statistic (see Table~3).

\subsection{Eastern Limb}\label{S-Limb}
\begin{figure}
	\includegraphics[width=\columnwidth]{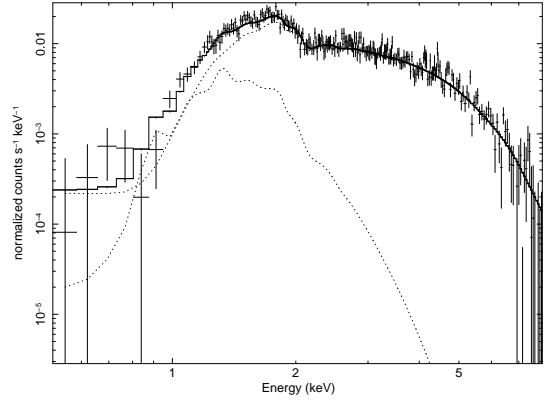}
	\caption{Combined Limb spectra plotted with the simultaneous fit to a power-law + \small{PSHOCK} model.}
	\label{FIG:Limb-fit}
\end{figure}

The nature of the extended emission surrounding the central PWN in G21.5$-$0.9 has been a puzzle since the first \textit{Chandra} observations (\cite{Slane-2000,Safi-Harb-2001}). Models including dust scattering and shock heated ejecta have been proposed. Imaging analysis by \cite{Matheson-2} found limb-brightening along the south eastern edge of the remnant, concluding that the dust scattering halo could not account for the total extended emission. Spectral analysis found an unreasonably high temperature for the emission to be purely thermal, which implied particle acceleration rather than shock-heated ejecta (Matheson \& Safi-Harb 2010). \cite{Nustar} found an excess of emission above 10 keV in the direction of the eastern limb with NuSTAR data, which further support the non-thermal model. Our results are presented in Table \ref{TAB:Limb-fit}, with a combined spectra shown in Figure \ref{FIG:Limb-fit}. The spectrum is best explained with a two-component power-law plus \small{PSHOCK} model. The emission is primarily non-thermal and the thermal component contributes only 3.5\% of the total flux. Although the thermal contribution is small, it is statistically required with a F-test probability of $2.7\times 10^{-11}$. No emission lines are observed and the thermal component is characterized by a temperature $kT=0.37$ keV and small ionization time-scale $n_{et} = 6.57 \times 10^{9}$ cm$^{-3}$ s (see Table \ref{TAB:Limb-fit}). For a limb emitting volume of $6\times 10^{56}D^{3}_{5}f cm^{3}$, where $f$ is the filling factor, the small amount of thermal emission suggests an emitting density of 1.76 cm$^{-3}$. The explosion energy implications of this density are further discussed in section \ref{SNR}.

\begin{table}%Skinny limb region
\caption{Spectral-fitting results for the eastern limb. See text for details.}
\label{TAB:Limb-fit}
\begin{tabular}{c c c}
\hline
\hline
Model & Parameter & Eastern Limb \\
\hline
& Effective Exposure & 318.7 ks \\
&  $N_{H} (10^{22}$ atoms cm$^{-2})$ & 3.237 (Frozen) \\
\hline
Power-law & $\Gamma$ & 2.49 (2.44-2.54) \\
& Norm ($10^{-4}$)$^{a}$ & 5.36 (5.11-5.63) \\
& Flux ($10^{-13}$)$^{b}$ & 5.38 (5.32-5.45) \\
& $ {\chi} ^{2}_{\nu}(\nu)$ & 1.016 (1040) \\
\hline
pshock & kT (keV) & 3.04 (2.85-3.37) \\
& $n_{et}$($10^{12}$ cm$^3$ s$^{-1}$) & 2.07 (0.97-4.78) \\
& Norm ($10^{-4}$ cm$^{-5}$) & 9.73 (8.87-10.02) \\
& $ {\chi} ^{2}_{\nu}(\nu)$ & 1.177 (1039) \\
& Flux ($10^{-13}$)$^{b}$ & 5.49 (5.42-5.58) \\
\hline
Power-law& $\Gamma$ & 2.22 (2.04-2.34) \\
+ & Norm ($10^{-4}$)$^{a}$ & 3.76 (2.90-4.41) \\
pshock & kT (keV) & 0.37 (0.20-0.64) \\
& $n_{et}$($10^{9}$ cm$^3$ s$^{-1}$) & 6.57 ($<29.5$) \\
& Norm ($10^{-3}$ cm$^{-5}$) & 6.21 (1.25-96.18) \\
& $ {\chi} ^{2}_{\nu}(\nu)$ & 0.966 (1037) \\
& Flux ($10^{-13}$)$^{b}$ & 5.61 (5.37-5.67) \\
& Non-thermal flux ($10^{-13}$)$^{b}$ & 5.41 \\
& Thermal flux ($10^{-13}$)$^{b}$ & 0.20 \\
\hline
Power-law$^{c}$& $\Gamma$ & 2.55 (Frozen) \\
+ & Norm ($10^{-4}$)$^{a}$ & 4.3 (0-4.9) \\
Power-law& $\Gamma$ & 1.5 (0.5-2.3) \\
+ & Norm ($10^{-5}$)$^{a}$ & 3.7 (0.6-44.6) \\
pshock & kT (keV) & 0.23 (0.15-0.37) \\
& $n_{et}$($10^{13}$ cm$^3$ s$^{-1}$) & 5 (Unconstrained) \\
& Norm ($10^{-3}$ cm$^{-5}$) & 4.2 (1.1-36.8) \\
& $ {\chi} ^{2}_{\nu}(\nu)$ & 0.966 (1036) \\
\hline
\multicolumn{3}{ l }{$^{a}$ Photons keV$^{-1}$ cm$^{-2}$ s$^{-1}$} \\
\multicolumn{3}{ l }{$^{b}$ Observed flux in erg cm$^{-2}$ s$^{-1}$}\\
\multicolumn{3}{ l }{$^{c}$ Dust scattering halo component}
\end{tabular}
\end{table}

%In a similar treatment to the northern knot, to try and decouple the scattered halo emission from the limb itself we extract a background from just inside the limb, however the limb brightening is not significant enough to provide usable statistics. 
In order to decouple the scattered halo emission from the limb itself, we extract a background from just inside the limb, with a similar treatment to the northern knot. However the limb brightening is not significant enough to provide robust statistics.
Instead we add a power-law component with the photon index fixed to the leveled off value from the radial profile and allow the normalization to vary. With this added non-thermal component we again find a faint thermal contribution with temperature of 0.23 keV (Table~4) - consistent within error of our non-halo result and a photon index of 1.5. Again, the added component does not improve the fitting statistic.

\section{Variability in the Pulsar Wind Nebula}
\begin{figure}
	\includegraphics[width=\columnwidth]{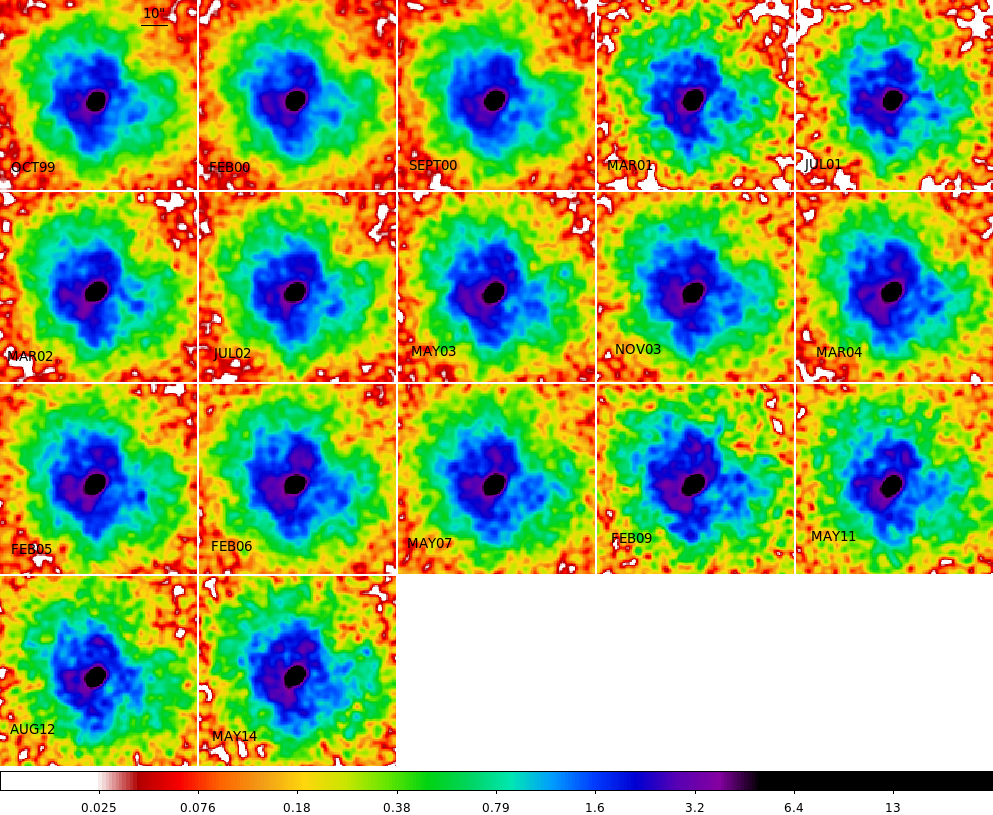}
	\caption{Variability seen with HRC.}
	\label{FIG:Variability-HRC}
\end{figure}

\begin{figure}
	\includegraphics[width=\columnwidth]{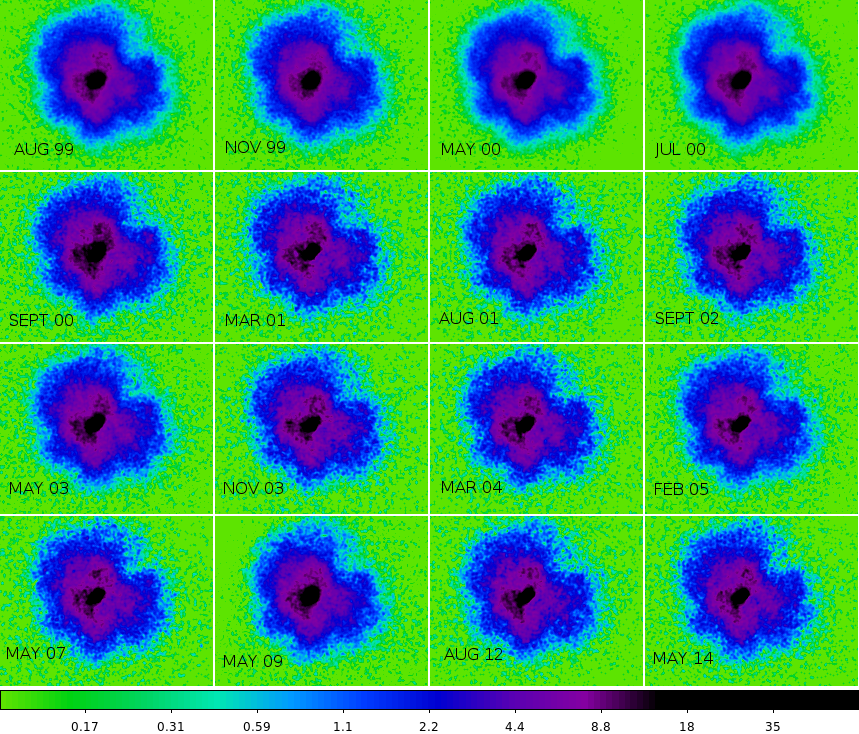}
	\caption{Variability seen with ACIS}
	\label{FIG:Variability-ACIS}
\end{figure}

\begin{figure}
	\includegraphics[width=\columnwidth]{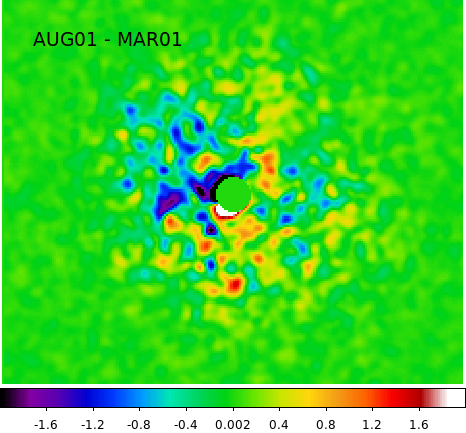}
	\caption{Sample ACIS difference image. The PSR has been removed to highlight the faint changes}
	\label{FIG:ACIS-Difference}
\end{figure}

Bright Knots, which appear and fade away on time-scales of weeks to months with velocities of $\sim 0.5c$, have been observed in PWNe such as the Crab and Vela nebulae \citep{Hester-Variability,Pavlov-Variability}. We searched for such features in G21.5$-$0.9 with the HRC observations from the same date, which were merged and normalized to 20~ks exposures. Figure \ref{FIG:Variability-HRC} shows the result. The same process was followed for the ACIS observations, which are displayed in Figure \ref{FIG:Variability-ACIS}. The images are available as a video slide show in the online journal. Unlike the Crab and Vela nebulae, G21.5$-$0.9 does not show persistent structure in the PWN.
To reveal changes, difference images (see Figure \ref{FIG:ACIS-Difference}) were created by subtracting one set of observations from the next.  

If we assume that the features we see are persistent structures which have moved rather than new wisps which formed between observations we can calculate the velocity required. Tracking several bright knots we find velocities of 0.2--0.75~c with an average of 0.5~c. Performing the same analysis on the ACIS observations yields consistent results.

%{\bf to include movies on-line, if permissible}

\section{Discussion}

\subsection{Pulsar J1833-1034}
The spectra of PSR J1833--1034 can be fitted reasonably well with a single power-law. The addition of a thermal blackbody component does improve the fit but only marginally. The F-test probability of 0.17 indicates that this component is unimportant compared to the previous result of \cite{Matheson-2}.%If we assume 
We assume this thermal component is real and then examine the effect of the fit on the pulsar parameters.  We assume black-body radiation, i.e., $L=4 \pi R^{2} \sigma T^{4}$. The unabsorbed thermal flux of $4.57\times 10^{-13}$ erg cm$^{-2}$ s$^{-1}$ at a distance of 5~kpc and temperature of 0.43~keV suggest an emitting region of 0.6~km in size, which is much smaller than the canonical radius of  a neutron star and is consistent with the previous suggestion that the source is a small hot spot on the neutron star surface. The spectral map of the PWN (Figure \ref{FIG:SpectralMapZoom}) reveals that the PSR is offset from the regions of harder spectra. If we attribute this offset to the termination shock radius, then the nebular magnetic field can be estimated as follows.  Given the spin-down energy loss $\dot{E}=3.37\times 10^{37}$ erg s$^{-1}$ \citep{Camilo2006}, $\theta_{s}=0.46'' \eta^{-1/2}d^{-1}_{5}B^{-1}_{mG}$, where $\theta_{s} = r_{s}/d$, $\eta$ is the filling factor of pulsar wind and $B_{mG}$ is the nebular magnetic field in mG. The offset of 3$^{\prime\prime}$.5 suggests a magnetic field of 0.13~mG.

%If we assume this thermal component is real, we examine the effect of the fit on the pulsar parameters.} We assume black-body radiation, i.e., $L=4 \pi R^{2} \sigma T^{4}$. The unabsorbed thermal flux of $4.57\times 10^{-13}$ erg cm$^{-2}$ s$^{-1}$ at a distance of 5~kpc and temperature of 0.43~keV suggest an emitting region of 0.6~km in size, which is much smaller than the canonical radius of  a neutron star and is consistent with the previous suggestion that the source is a small hot spot on the neutron star surface. The spectral map of the PWN (Figure \ref{FIG:SpectralMapZoom}) reveals that the PSR is offset from the regions of harder spectra. If we attribute this offset to the termination shock radius, then the nebular magnetic field can be estimated as follows.  Given the spin-down energy loss $\dot{E}=3.37\times 10^{37}$ erg s$^{-1}$ \citep{Camilo2006}, $\theta_{s}=0.46'' \eta^{-1/2}d^{-1}_{5}B^{-1}_{mG}$, where $\theta_{s} = r_{s}/d$, $\eta$ is the filling factor of pulsar wind and $B_{mG}$ is the nebular magnetic field in mG. The offset of 3$^{\prime\prime}$.5 suggests a magnetic field of 0.13~mG.

\subsection{Pulsar wind nebula}{\label{sec:PWN_fitting}}
 \cite{KC84a,KC84b} (hereafter KC) treat the pulsar wind as magnetohydrodynamic (MHD) flow and then derive the steady-state particle and magnetic field structure in the PWN with spherical symmetry and a purely advective wind flow. The KC model was able to explain the spectral and spatial distribution of the optical and X-ray emission in the Crab Nebula, but failed in the radio band. The discovery of axisymmetric jet-torus structures in PWNe like the Crab nebula with high resolution X-ray and optical imaging \citep[e.g.][]{Weisskopf00,hester08} motivates 2-dimensional (2D) MHD simulations of PWNe with anisotropic pulsar wind power. Current 2D MHD simulations are able to reproduce the jet-torus structure and the inner ring feature in PWNe with polar angle dependent pulsar wind power \citep[e.g.][]{KL03,Del06}. However, toroidal structures are only detected in the inner part of the nebula \citep[e.g.][]{Safi-Harb-2001,Slane04,hester08}, where the impact of the pulsar wind is crucial. In the outer part of the nebula, complex filamentary structures with fingers and loops are instead observed \citep[e.g.][]{Seward06,hester08}. It is found that the standard MHD model ran into problems to explain the spectral index distribution and surface brightness profile in the outer part of the nebula \citep{Amato00,Slane04}. The polarization measurements \citep[e.g.][]{hester08} and the deep \textit{Chandra} images \citep[e.g.][]{Seward06} indicate that the magnetic topology in the filamentary structure is much more complicated than the toroidal field assumed in the standard MHD model. 

\cite{TC12} show that the introduction of diffusive particle transport in PWNe can explain the spectral index distribution and the nebular size behavior in young PWNe like the Crab and 3C~58 very well. It is assumed that advection plays a dominant role in the inner part of the nebula with toroidal structure, while diffusion becomes dominant in the outer part of the nebula with filamentary structure. The exact nature of diffusion is still unclear, and is likely induced by the Rayleigh-Taylor (RT) instability at the outer boundary of the nebula \citep[e.g.][]{CG75,Jun98} or/and the kink instability triggered at the TS \citep[e.g.][]{Begelman98,Camus09}. These fluid instabilities may be able to destroy the ordered toroidal field, which is imposed by the pulsar wind, and drive turbulence in the PWN. Recent 3-dimensional MHD and test particle simulations reveal a turbulent nebula with high velocity fluctuation \citep{Porth16}, which indicates efficient diffusive transport of particles in PWNe. The effective diffusion coefficient is estimated as \citep{Porth16}.

\begin{equation}
D_{eff}\sim 2\times 10^{27}~ \left( \frac{L_{TS}}{0.13\rm pc}\right)~\rm cm^2 s^{-1}
\label{eq:diffusion}
\end{equation}
which is found to be independent of energy for electrons up to PeV energy. $L_{TS}$ is the radius of TS.

\begin{figure}
        \center{\includegraphics[width=\columnwidth] {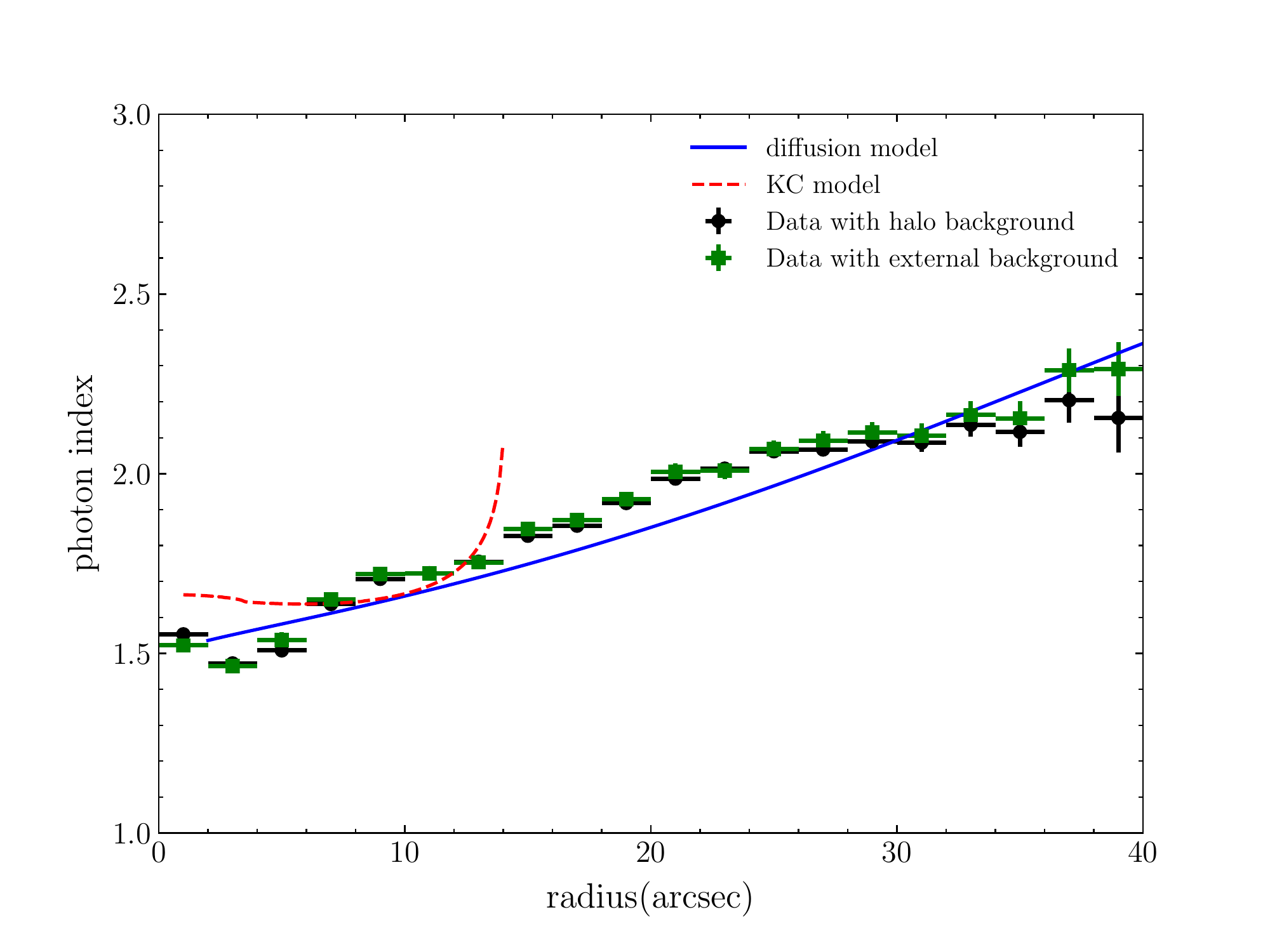}}
        \caption{\label{fig:PWN_fitting} Fitting of the photon index distribution between $0.5$~keV and $8$~keV in G21.5--0.9. The black and green crossings are the data with halo and external background respectively. The blue solid line is the pure diffusion model fitting with $\alpha=1.9$, $B = 130 \mu G$ and $D \sim 2.1\times 10^{27}\rm cm^2/s$. The red dashed line is the KC model fitting with $\alpha=1.9$, $B_{TS} = 50 \mu G$  and $\sigma=3\times 10^{-3}$.
}
\end{figure}

In Fig. \ref{fig:PWN_fitting}, we fit the spectral index distribution between $0.5$~keV and $8$~keV in G21.5-0.9 with both the KC model and a pure diffusion model \citep{TC12} assuming simple spherical symmetry. The red dashed line represents the KC model results with $\sigma=3\times 10^{-3}$, $\alpha=1.9$ and $B_{TS}=50 \mu G,$\footnote{In the KC model, the flow velocity decreases very quickly with radius in the postshock region. As a result, a small magnetic field is needed to explain the observed large nebular size. However, even with a small magnetic field, the KC model cannot reproduce the spectral index profile successfully as shown in Fig. \ref{fig:PWN_fitting}, which motivates the introduction of diffusive transport of particles.}
 where $\sigma$ is the Poynting flux to particle energy flux ratio at the TS, $\alpha$ is the power law index of the injected particle spectrum and $B_{TS}$ is the magnetic field at the TS. $\sigma$ is chosen to be $3\times 10^{-3}$ to match the expansion velocity of  $910$ km/s at the outer boundary of G21.5--0.9 \citep{Radio-Age}. It is clear that the KC model is inconsistent with the spectral index distribution in G21.5--0.9, which is similar to the previous study of 3C~58 \citep{Slane04}. The blue solid line presents the diffusion model results with $\alpha=1.9$, $D \sim 2.1\times 10^{27}\rm cm^2/s$ and $B =130 \mu G$ (see \S5.1). $D$ and $B$ are the diffusion coefficient and magnetic field respectively, which are assumed to be constant for simplification and should be understood as the spatial averaged values in the PWN. The angular radius of the TS is assumed to be $\theta_{s} \sim$3$^{\prime\prime}$.5. It is interesting to note that the effective diffusion coefficient $D_{eff}$ defined in eq. (\ref{eq:diffusion}) is estimated to be $\sim 1.3 \times 10^{27}\rm cm^2/s$ with $L_{TS}\sim 0.08$pc, which agrees with our diffusion model results.

In the pure diffusion model, the photon index distribution of the nebula is determined by the dimensionless ratio $\zeta=r^2/Dt_c$, where $r$ is the radius and $t_c$ is the cooling time scale. If $\zeta\ll 1$, the photon index distribution is very flat. If $\zeta\gtrsim 1$, the photon index distribution gradually steepens as $\zeta$ increases. For synchrotron-dominated cooling, $\zeta \propto r^2B^{3/2}\nu^{1/2}/D$, where $\nu$ is the corresponding emission frequency. 
The diffusion coefficient of charged particles in a nebula
depends on the magnetic field configuration and the Larmor radius.
The nature of this magnetic field dependence is not fully understood yet. In the limit of Bohm diffusion, $D\propto B^{-1}$ and $\zeta \propto r^2 B^{5/2}\nu^{1/2}$ which is considered to be a reasonable choice in the presence of strong turbulence \citep[e.g.,][]{HS14}. Field line random walk represents another limiting case, in which magnetic field lines wander due to turbulence and the particles follow the field lines exclusively. In this case, the diffusion coefficient scales as $D\propto B^{-2}$ and $\zeta \propto r^2 B^{7/2}\nu^{1/2}$ \citep[e.g.,][]{Shalchi09}. The magnetic dependence of $D$ in the nebula is likely between the two limiting cases.

The spatial dependence of the magnetic field in the nebula is very complicated according to 2D and 3D MHD simulations \citep{Del06,Porth16}, which is beyond the scope of this work. Here we briefly discuss the magnetic configuration in the KC model. Behind the termination shock, the magnetic field  evolves as $B\propto r $ due to the flux freezing. As $r$ increases, the magnetic pressure gradually becomes dominant. After that, the radial speed is approximately constant and $B \propto r^{-1}$ instead. In the $B\propto r$ regime, which corresponds to the inner part of the nebula, we have $\zeta \propto r^{9/2}$ for Bohm diffusion and $\zeta \propto r^{11/2}$ for field line random walk respectively. If $\zeta $ increases with radius, we expect the photon index distribution steepens much faster with radius compared to our simplified calculation with constant $D$ and $B$. In the $B\propto r^{-1}$ regime, which corresponds to the outer part of the nebula, we instead have $\zeta \propto r^{-1/2}$  and $\zeta \propto r^{-3/2}$ for the two limiting cases of diffusion. 
If $\zeta $ decreases with radius, the photon index distribution  becomes more flat compared to our simplified calculation. In summary, if we consider the spatial dependence of $B$ and $D$ described above, then the photon index distribution steepens in the inner region and flattens in the outer region, which appears to be more consistent with the data of G21.5--0.9.

Based on the above discussion, when $D\propto B^{3/2}$, $\zeta$ is the same and the photon index distribution remains almost the same. If we instead assume $B = 50 \mu G$ as indicated by \cite{De08}, then we obtain $D \sim 5\times 10^{26}\rm cm^2/s$, which is consistent with the results derived in \cite{Porth16} based on 3D MHD simulations. Recently, G21.5--0.9 was discussed by \cite{Lu17} with an improved model including dynamical evolution of the central pulsar, energy dependence of diffusion coefficient $D$, and radial dependence of diffusion coefficient $D$ and magnetic field $B$. It is not straightforward to compare with their fitting results directly. Here we only consider the spatially averaged value derived at present day. \cite{Lu17} found that $B\sim 40 \mu G$ and $D\sim 1\times 10^{26}cm^2s^{-1}$ for the energy range of interest here, which is consistent with our simple toy model results surprisingly well (we note that the diffusion coefficient shown in Table~2 of \cite{Lu17} is at a much lower electron energy). According to the above discussion, if we assume the diffusion coefficient provided in eq (\ref{eq:diffusion}) is a good approximation, then we can decouple $D$ and $B$ in the pure diffusion model and apply the model to estimate the spatially averaged magnetic field in a PWN.

\subsection{Supernova Remnant}\label{SNR}
The absence of a shell-like structure surrounding some PWNe is a long standing puzzle. Whether the undetected shells are due to expansion into a very low density medium or a consequence of a low energy electron-capture supernova explosion (e.g., \cite{Nomoto-1982,YC15}) remains an open question. 
%Add: http://adsabs.harvard.edu/abs/2015ApJ...806..153Y
Deep observations revealed thermal emission in 3C~58 \citep{3c58-2001,3c58-Thermal} and G54.1+0.3 \citep{g54}, which is attributed to the missing shells and/or shock-heated supernova ejecta. The faint thermal emission detected in the eastern limb may be used to calculate the shock speed and estimate the energy released during the supernova event. The Rankine-Hugoniot relations for an adiabatic shock \citep{SNR-HighEnergy} yield $kT=\frac{3}{16}\mu m_{p}v^{2}_{s}$, where $\mu$ is the mean mass per particle ($\mu \approx 0.6$), $m_{p}$ is the proton mass and $v_{s}$ is the shock speed. The observed temperature of 0.37~keV corresponds to a shock speed of 562~km/s. Assuming a constant shock velocity and 5~kpc distance, the 140$^{\prime\prime}$ shell radius gives an age of 1.2~kyr, which is slightly larger than the 870 years derived from radio observations of the nebula expansion with a velocity of 910 km/s. For young remnants with an age less than 1000 years, the forward shock speed is expected to be higher. Although the derived temperature is low, it is consistent with that found in 3C~58 \citep{3c58-Thermal,3c58-2001}, which is described by a dominant non-thermal component with $\Gamma \sim 2-3$ and a weaker thermal component with $kT \sim 0.2$ keV. However, 3C58 is older, and more evolved, with the thermal emission residing at the outer boundary of the PWN rather than a distinct and separate limb-brightened component. The low temperature may indicate a non-equilibrium state of the electrons and ions as suggested by \cite{Electron-ion-heating}, therefore we note that the estimated shock velocity scales as $\sqrt{T_{p}/T_{e}}$.
We use the thermal component to calculate the ambient density into which the shock wave is expanding. For a limb emitting volume of $6\times 10^{56}D^{3}_{5}f cm^{3}$, where $f$ is the filling factor, we find a number density of $1.76~cm^{-3}$ which corresponds to a limb mass of $\sim 1 M_\odot$. If we assume this is the swept-up mass, then the remnant must be expanding into a low-density medium with $n_e \sim 0.19~cm^{-3}$, which is smaller than the upper limit of $0.65~cm^{-3}$ provided by \cite{Bocchino-2005}. Extending the limb to a full sphere, we estimate the kinetic energy of the supernova to be $3\times 10^{49}$ erg, which is much smaller than the $10^{51}$~erg expected for a typical supernova. The minimal thermal emission may indicate the SNR is expanding into the wind blown bubble produced by its progenitor. \cite{SNR-SizeDistribution} predicts cavities with radii of $\approx$ 2--20~pc. G21.5$-$0.9 is consistent with the lower limit of this range. The low mass loss prior to a type IIP supernova combined with the wind blown cavity suggests the shock has not yet swept up enough mass to transition into a Sedov-Taylor phase.

Future observations with a sensitive, high-resolution, spectrometer such as the X-Ray Imaging and Spectroscopy Mission ($XRISM$, formerly known as $XARM$) in the near future, and $ATHENA$ in the more distant future, should reveal the missing thermal X-ray emission from the shocked ambient or circumstellar material in G21.5--0.9 and other shell-less PWNe.
% \section{Conclusions}

% The last numbered section should briefly summarise what has been done, and describe
% the final conclusions which the authors draw from their work.

\section*{Acknowledgements}
%
%The Acknowledgements section is not numbered. Here you can thank helpful
% colleagues, acknowledge funding agencies, telescopes and facilities used etc.

We thank Roger Chevalier for comments on the manuscript. This research made use of NASA's Astrophysics Data System and the HEASARC operated by NASA's Goddard Space Flight Center. S.S.H. acknowledges support by NSERC through the Discovery Grants and the Canada Research Chairs programs, and by the Canadian Space Agency. 
B.G. acknowledges support from a University of Manitoba Graduate Fellowship.
We thank the referee for their careful reading of the paper which helped improve its quality and clarity.
%%%%%%%%%%%%%%%%%%%%%%%%%%%%%%%%%%%%%%%%%%%%%%%%%%



\bibliographystyle{mnras}
\bibliography{G21Bib} % if your bibtex file is called example.bib

% \begin{thebibliography}{99}
% \bibitem[\protect\citeauthoryear{Author}{2012}]{Author2012}
% Author A.~N., 2013, Journal of Improbable Astronomy, 1, 1
% \bibitem[\protect\citeauthoryear{Others}{2013}]{Others2013}
% Others S., 2012, Journal of Interesting Stuff, 17, 198
% \end{thebibliography}

%%%%%%%%%%%%%%%%%%%%%%%%%%%%%%%%%%%%%%%%%%%%%%%%%%

%%%%%%%%%%%%%%%%% APPENDICES %%%%%%%%%%%%%%%%%%%%%

%{\bf need to refer to the Appendices in the paper}

\appendix

\section{PWN Radial Profile data}
% If you want to present additional material which would interrupt the flow of the main paper,
% it can be placed in an Appendix which appears after the list of references.
\begin{table}
\caption{Spectral fit data found for the 5--8 keV radial profile with a background taken from inside the halo. The absorption model \small{TBABS} is frozen at 3.222$\times$10$^{22}$~cm$^{-2}$. Errors quoted are 90\% confidence levels.}
\label{TAB:HaloBackground-5-8}
\begin{tabular}{c c c c c}
\hline\hline
Radius	&	$\Gamma$	&	error	&	Norm ($10^{-4}$)	& error ($10^{-4}$)\\
(arcsec) & & $\pm$& &$\pm$\\
\hline
0-2	&	1.28	&	0.14	&	6	&	2	\\
2-4	&	1.48	&	0.13	&	9	&	2	\\
4-6	&	1.55	&	0.18	&	7	&	3	\\
6-8	&	1.60		&	0.15	&	8	&	3	\\
8-10	&	1.68	&	0.15	&	11	&	3	\\
10-12	&	1.76	&	0.15	&	14	&	4	\\
12-14	&	1.93	&	0.15	&	17	&	5	\\
14-16	&	1.86	&	0.15	&	14	&	5	\\
16-18	&	1.75	&	0.18	&	10	&	4	\\
18-20	&	1.97	&	0.20	&	13	&	5	\\
20-22	&	1.96	&	0.23	&	10	&	5	\\
22-24	&	2.08	&	0.26	&	11	&	6	\\
24-26	&	2.28	&	0.30	&	13	&	9	\\
26-28	&	2.28	&	0.31	&	12	&	9	\\
28-30	&	2.36	&	0.36	&	11	&	10	\\
30-40	&	2.16	&	0.26	&	16	&	10	\\

\hline

\end{tabular}
\end{table}

\begin{table}
\caption{Spectral fit data for the whole SNR in the 0.5--8 keV radial profile with a background taken from inside the halo. The absorption model \small{TBABS} is frozen at 3.222$\times$10$^{22}$~cm$^{-2}$. Errors quoted are 90\% confidence levels.}
\label{TAB:HaloBackgaround-05-8}
\begin{tabular}{c c c c c}
\hline\hline
Radius 	&	$\Gamma$	&	error &	Norm ($10^{-4}$)	& error ($10^{-4}$)\\
(arcsec) & &$\pm$ & & $\pm$\\
\hline
0-2	&	1.55	&	0.01	&	10.1	&	0.2	\\
2-4	&	1.47	&	0.01	&	9.2		&	0.1	\\
4-6	&	1.51	&	0.01	&	6.8		&	0.1	\\
6-8	&	1.64	&	0.02	&	8.4		&	0.1	\\
8-10	&	1.71	&	0.01	&	11.3	&	0.2	\\
10-12	&	1.72	&	0.01	&	13	&	0.2	\\
12-14	&	1.76	&	0.01	&	13.2	&	0.2	\\
14-16	&	1.83	&	0.01	&	13.3	&	0.2	\\
16-18	&	1.85	&	0.01	&	12.5	&	0.2	\\
18-20	&	1.92	&	0.02	&	11.5	&	0.2	\\
20-22	&	1.99	&	0.02	&	10.6	&	0.2	\\
22-24	&	2.01	&	0.02	&	9.7	&	0.2	\\
24-26	&	2.06	&	0.02	&	9.1	&	0.2	\\
26-28	&	2.07	&	0.02	&	8.3	&	0.2	\\
28-30	&	2.09	&	0.02	&	6.9	&	0.2	\\
30-32	&	2.09	&	0.03	&	5.2	&	0.2	\\
32-34	&	2.14	&	0.03	&	4	&	0.1	\\
34-36	&	2.12	&	0.04	&	2.7	&	0.1	\\
36-38	&	2.20	&	0.06	&	1.8	&	0.1	\\
38-40	&	2.16	&	0.1	&	1	&	0.1	\\

\hline
\end{tabular}
\end{table}

\begin{table}
\caption{Spectral fit data found for the whole SNR in the 0.5--8 keV radial profile with a background taken from outside the observed SNR shell. The absorption model \small{TBABS} is frozen at 3.237$\times$10$^{22}$~cm$^{-2}$. Errors quoted are 90\% confidence levels.}
\begin{tabular}{c c c c c}
\hline\hline
Radius	&	$\Gamma$	&	error	&	Norm ($10^{-4}$)	& error ($10^{-4}$)\\
(arcsec) & & $\pm$& &$\pm$\\
0-2	&	1.52	&	0.02	&	9.5	&	0.2	\\
2-4	&	1.46	&	0.02	&	8.9	&	0.2	\\
4-6	&	1.54	&	0.02	&	7.0	&	0.2	\\
6-8	&	1.65	&	0.02	&	8.6	&	0.2	\\
8-10	&	1.72	&	0.02	&	11.7	&	0.3	\\
10-12	&	1.72	&	0.02	&	13.3	&	0.3	\\
12-14	&	1.75	&	0.02	&	13.4	&	0.3	\\
14-16	&	1.85	&	0.02	&	13.9	&	0.3	\\
16-18	&	1.87	&	0.02	&	13.0	&	0.3	\\
18-20	&	1.93	&	0.02	&	12.1	&	0.3	\\
20-22	&	2.01	&	0.02	&	11.2	&	0.3	\\
22-24	&	2.01	&	0.02	&	10.1	&	0.3	\\
24-26	&	2.07	&	0.03	&	9.8	&	0.3	\\
26-28	&	2.09	&	0.03	&	9.2	&	0.3	\\
28-30	&	2.11	&	0.03	&	7.7	&	0.3	\\
30-32	&	2.11	&	0.03	&	6.0	&	0.2	\\
32-34	&	2.16	&	0.04	&	4.9	&	0.2	\\
34-36	&	2.15	&	0.05	&	3.6	&	0.2	\\
36-38	&	2.29	&	0.06	&	2.8	&	0.2	\\
38-40	&	2.29	&	0.08	&	2.1	&	0.2	\\
40-44	&	2.37	&	0.06	&	3.1	&	0.2	\\
44-48	&	2.46	&	0.08	&	2.5	&	0.2	\\
48-52	&	2.58	&	0.08	&	2.5	&	0.2	\\
52-56	&	2.60	&	0.08	&	2.5	&	0.2	\\
56-60	&	2.56	&	0.08	&	2.4	&	0.2	\\
60-64	&	2.62	&	0.09	&	2.6	&	0.2	\\
64-68	&	2.60	&	0.09	&	2.3	&	0.2	\\
68-72	&	2.61	&	0.08	&	2.7	&	0.2	\\
72-76	&	2.56	&	0.08	&	2.9	&	0.2	\\
76-80	&	2.66	&	0.07	&	3.7	&	0.3	\\
80-84	&	2.65	&	0.07	&	3.9	&	0.3	\\
84-88	&	2.49	&	0.07	&	3.1	&	0.2	\\
88-92	&	2.48	&	0.08	&	2.7	&	0.2	\\
92-96	&	2.59	&	0.08	&	2.9	&	0.2	\\
96-100	&	2.53	&	0.07	&	2.8	&	0.2	\\
100-104	&	2.49	&	0.07	&	2.8	&	0.2	\\
104-108	&	2.58	&	0.08	&	3.0	&	0.2	\\
108-112	&	2.52	&	0.08	&	2.6	&	0.2	\\
112-116	&	2.64	&	0.07	&	4.1	&	0.3	\\
116-120	&	2.51	&	0.08	&	2.5	&	0.2	\\
120-124	&	2.58	&	0.08	&	2.5	&	0.2	\\
124-128	&	2.55	&	0.09	&	2.3	&	0.2	\\
128-132	&	2.53	&	0.09	&	2.3	&	0.2	\\
132-136	&	2.49	&	0.09	&	2.4	&	0.2	\\
136-140	&	2.36	&	0.08	&	2.0	&	0.2	\\
140-144	&	2.46	&	0.10	&	1.8	&	0.2	\\
144-148	&	2.53	&	0.11	&	1.6	&	0.2	\\
148-152	&	2.66	&	0.13	&	1.6	&	0.2	\\
152-156	&	2.65	&	0.17	&	1.2	&	0.2	\\
156-160	&	2.96	&	0.20	&	1.2	&	0.2	\\

\hline
\end{tabular}
\end{table}

\begin{table}
\caption{Spectral fit data for the dust scattering halo. Spectra were extracted from annuli segments spanning 280-360 degrees  where the SNR does not display limb brightening or knot features and beginning outside the edge of the PWN. The absorption model TBABS is frozen at $3.237\times 10^{22}cm^{-2}$. Errors quoted are 90\% confidence levels.}
\begin{tabular}{c c c c c}
\hline\hline
Radius	&	$\Gamma$	&	error	&	Norm ($10^{-4}$)	& error ($10^{-4}$)\\
(arcsec) & & $\pm$& &$\pm$\\
45-55	&	2.47	&	0.12	&	1.4	&	0.2	\\
55-65	&	2.49	&	0.15	&	1.1	&	0.2	\\
65-75	&	2.40	&	0.17	&	1.0	&	0.2	\\
75-85	&	2.63	&	0.17	&	1.1	&	0.2	\\
85-95	&	2.54	&	0.17	&	1.0	&	0.2	\\
95-105	&	2.64	&	0.17	&	1.1	&	0.2	\\
105-115	&	2.54	&	0.19	&	0.8	&	0.2	\\
115-125	&	2.96	&	0.21	&	1.0	&	0.2	\\
125-135	&	2.52	&	0.20	&	0.8	&	0.2	\\
135-145	&	2.58	&	0.18	&	1.0	&	0.2	\\
145-155	&	2.48	&	0.19	&	0.8	&	0.2	\\
155-165	&	2.60	&	0.30	&	0.5	&	0.2	\\
\hline
\end{tabular}
\end{table}

%%%%%%%%%%%%%%%%%%%%%%%%%%%%%%%%%%%%%%%%%%%%%%%%%%

% Don't change these lines
\bsp	% typesetting comment
\label{lastpage}
\end{document}